\documentstyle[aps]{revtex}
\input epsf.tex

\begin{document}
\title{Time variability of the gravitational constant and Type Ia supernovae}
\author{Luca Amendola, Stefano Corasaniti, Franco Occhionero}
\address{Osservatorio Astronomico di Roma, \\
Viale Parco Mellini 84, \\
00136 Roma, Italy}
\date{\today }
\maketitle

\begin{abstract}
We investigate to which extent a time variation of the gravitational
constant or other fundamental constants affects the best fit of the Hubble
diagram of type Ia supernovae. In particular, we show that a slow increase
of $G$ in the past, below experimental constraints, can reconcile the SNIa
observations with an open zero-$\Lambda $ universe.
\end{abstract}

\section{Introduction}

Type Ia supernovae are an exciting tool for cosmology. They are likely to be
good standard candles, and can be observed down to $z\sim 1$. Residual
systematic effects, like reddening, $K$-correction, and intrinsic
differences in the light curve, can be removed by suitable techniques, so
that their intrinsic peak magnitude can be estimated with an error of less
than half a magnitude down to $m\sim 24$, where different cosmological
models begin to be distinguishable. Recent work by Perlmutter et al. (1999,
hereinafter P99) and Riess et al. (1998) have shown that SNe Ia are indeed a
powerful test of world models. The most striking results so far, obtained
with 42 deep redshift SNe, and 18 low redshift SNe (P99; see also Garnavich
et al. 1998) are that a zero-$\Lambda $ flat universe is ruled out to an
extremely high confidence level (c.l.), and that the best-fit flat universe
is $\Omega _{m}=0.28$ and $\Omega _{\Lambda }=0.72$. An open, zero-$\Lambda $
model is also ruled out, at 99\% c.l.. These results are particularly
important because an $\Omega _{m}$ around 0.3 is in agreement with a number
of independent observations, especially concerning the mass of clusters.

The observed SNe at high redshift are roughly half a magnitude fainter than
what a flat matter-dominated universe predicts. Such a relatively small
difference, and the crucial result that depends on it, deserves close
examination (see for instance an alternative physical explanation in Goodwin
et al. 1999, and a phenomenological one in Drell et al. 1999). In this paper
we investigate whether a slow time dependence of some fundamental constant
can affect the result. The reason for such an investigation is intuitive: if
the peak luminosity $L_{p}^{\text{ }}$ of SNe Ia depends on, say, the
gravitational constant $G$, then a relation $G(z)$ will show up as a $%
L_{p}(z),$ thereby modifying the assumption of standard candles. In
particular, suppose $G$ {\it increases} by $\Delta G$ at a certain look-back
time $t$. Suppose then that $L_{p}\sim G^{-\gamma },\gamma >0$ (see below);
then the predicted apparent magnitude will be fainter by 
\begin{equation}
\Delta m_{G}=2.5\gamma \frac{\Delta G(t)}{\left( \ln 10\right) G}.
\end{equation}
Let us call such an effect $G$-correction. If $\gamma $ is of order unity,
and $G$ at $z\sim 0.5$, where most deep redshift SNe are, was, say, 50\%
higher than the present value, then the $G$-correction would be half a
magnitude, and could reconcile the SN observations with a zero-$\Lambda $
cosmology. Naturally, we are thinking that $G$ increases monotonically with
look-back time, so that the low redshift SNe, as well as the light curve
calibration, remain unaffected by the $G$-correction.

SNe Ia are expected to be standard candles because the explosion occurs when
a white dwarf accretes matter from a companion and reaches the Chandrasekhar
mass 
\begin{equation}
M_{C}\simeq \frac{3.1}{m^{\prime }}\left( \frac{\hbar c}{G}\right) ^{3/2},
\end{equation}
where $\ m^{\prime }$ is the mass per electron, independently of the
progenitor status and of the accretion history (at least in the standard
model, see e.g. Woosley \& Weaver 1986). It is then very likely that some
relation between the peak luminosity and the Chandrasekhar mass exists,
whatever the precise explosion mechanism is (see e.g. Arnett 1982). It seems
therefore reasonable to test the hypothesis 
\begin{equation}
L_{p}\sim G^{-\gamma },
\end{equation}
with $\gamma >0$ of order unity.

Actually, any dependence of $L_{p}$ on a fundamental constant, like the
nucleon mass, can give an effect similar to the $G$-correction. In fact, a
field theory with a coupling of a Brans-Dicke field to gravity can always be
transformed into a mathematically equivalent theory in which the field
couples explicitely to the matter fields, rather than to gravity, and
therefore the masses are field-dependent. In the following, for
definiteness, we focus however on the gravitational constant.

\section{The $G$-correction}

Let us then formulate more precisely the effect we are testing. There are
several models which predict a time dependence of $G$, generally based on a
Brans-Dicke coupling. Most of them can be simply parameterized as 
\begin{equation}
G=G_{0}(1+tH_{0})^{m},  \label{pl}
\end{equation}
where $G_{0}$ is the present gravitational constant, $t$ is the look-back
time, and $H_{0}\simeq 10^{-10}yr^{-1}$ is the present Hubble length for $h=1
$ (see e.g. Amendola 1999a). The constraints on $\dot{G}$ can be expressed
as follows 
\begin{equation}
|\frac{\dot{G}}{G}|_{0}=|m|H_{0}<a10^{-10}yr^{-1}.
\end{equation}
Therefore we can assume 
\begin{equation}
|m|<a.
\end{equation}
The value of $a$ is of the order of 0.1-0.01, depending on the assumptions
(see e.g. Guenther et al. 1996). The relation between the look-back time $%
\tau $ in units of $H_{0}^{-1}$ and the redshift is 
\begin{equation}
\tau \equiv tH_{0}=\int_{0}^{z}\frac{dz^{\prime }}{1+z^{\prime }}\left[
(1+z^{\prime })^{2}(1+\Omega _{m}z^{\prime })-z^{\prime }(2+z^{\prime
})\Omega _{\Lambda }\right] ^{-1/2}.  \label{tz}
\end{equation}
We can write then $L_{p}=L_{p0}(G/G_{0})^{-\gamma }$ and obtain a $G$%
-correction 
\begin{equation}
\Delta m_{G}=2.5\beta \log [1+\tau (z)],
\end{equation}
where $\beta =m\gamma $. The maximum value of $\beta $ is of course crucial
to our argument. However, it is difficult to determine it with some
certainty, especially because the SNe Ia mechanism is still matter of
controversy. It is probably safe to assume $\gamma \simeq 1$, which implies $%
\beta <0.1-0.01$. If the stronger constraint on $m$ holds, then the $G$%
-correction is probably unable to alter the conclusions of P99 and Garnavich
et al. (1998). However, there are at least two possibilities to weaken the
constraints on $\beta $. First, the dependence of $L_{p}$ on $G$ can be
stronger than $\gamma \simeq 1$. Second, the $G(t)$ relation can be
different from the simple power-law (\ref{pl}) : for instance, $G$ could be
an oscillating function of time, as was proposed by Morikawa (1991) to
explain the periodicity in the pencil-beam galaxy distribution. In this
case, tuning the oscillating phase and choosing a period of the order of $%
z\simeq 0.5-0.8$, one can have 
\mbox{$\vert$}%
$\dot{G}/G|$ very small at present, and larger in the past, so as to escape
the present constraints. An oscillating $G(t)$ would be naturally produced
if the Brans-Dicke scalar field oscillates around its potential minimum. For
such a behavior, our power law approximation is supposed to hold
approximately around $z\simeq 0.5-0.8$.

For these reasons, and because in this paper we are interested in testing to
what extent a time dependence of the fundamental constants affects the SN
results, we allow $\beta $ to vary even beyond 0.1. Actually, the SNe Ia
measures can be employed also to estimate $\beta $ itself, as an additional
parameter to $\Omega _{m},\Omega _{\Lambda }$, as we will show.

There is actually a small inconsistency with Eq. (\ref{tz}) and in the
luminosity distance: since we are assuming $G$ to be time-dependent, we
should use the Brans-Dicke equation, or some variants of a scalar-tensor
theory, in place of the Friedmann equation. Or, equivalently, if we leave $G$
constant, and vary one of the other fundamental constants, say the nucleon
mass, we should insert an explicit coupling between matter and the scalar
field, and this would change the dependence of the matter density with time
in the Friedmann equation (see e.g. Wetterich 1995, Amendola 1999b).
However, the tight constraint on $\dot{G}/G$ implies that the deviation from
the Friedmann equation is minimal, so that we make only a small error in
using it. There is also the possibility that the non-minimally coupled field
is itself a dynamical cosmological constant, as in quintessence models
(Caldwell et al. 1998, Uzan 1999, Chen \& Kamionkowsky 1999, Baccigalupi et
al. 1999, Amendola 1999c). In these models, a new parameter is needed, the
scalar field effective equation of state $w=p/\rho $, and the calculations
should take into account explicitely this. Such models will be investigated
in another paper.

\section{Likelihood results}

Once the $G$-correction is added to the expected apparent magnitude, we
obtain the following relation 
\begin{equation}
m(z)={\cal M}+5\log {\cal D}_{L}(z;\Omega _{m},\Omega _{\Lambda })+\Delta
m_{G}(z;\Omega _{m},\Omega _{\Lambda },\beta ),
\end{equation}
where we use the notation of P99, in which ${\cal M}$ is the
Hubble-constant-free magnitude zero-point, and ${\cal D}_{L}$ the
Hubble-constant-free luminosity distance. In Fig. 1 we show $m(z)$ for
various choices of the parameters, fixing ${\cal M}$ to its best fit value.
Values of $\beta $ of the order of 0.5 or larger seem to explain the data
even for $\Lambda =0$. To be quantitative, we form as in P99 a gaussian
likelihood function $L({\cal M},\Omega _{m},\Omega _{\Lambda },\beta )$ for
the four parameters to be estimated. With respect to P99 we adopt two
simplifications. First, we neglect the dependence on a fifth parameter,
called $\alpha $ in P99, the slope of the width-luminosity relation, and
assume throughout the P99 best value, $\alpha =0.6$. As stated in P99, the
dependence on $\alpha $ has anyway a small effect on the Hubble diagram. The
result we get for $\beta =0$, extremely close to the original P99 results,
confirm that this is indeed an acceptable approximation. The second
approximation is to neglect the correlations among the photometric
uncertainties for the high-redshift SNe; as explained in P99, they are small
and, again, the comparison between our results and the original ones
confirms that this approximation is not harmful.

As in P99, we marginalize the likelihood function over the parameter ${\cal M%
}$ (that is, we integrate it over), we assume the prior condition $\Omega
_{m}>0$ and, finally, we use the catalog of SNe Ia published in P99,
excluding the six SNe rejected for the fit C. We are left then with 16
low-redshift and 38 high-redshift SNe.

We first study how the confidence regions for $\Omega _{m},\Omega _{\Lambda }
$ depend on $\beta .$ In Fig. 2 and 3 we show the confidence regions for $%
\beta =0,0.2$ and $0.5$. For $\beta =0$ we recover the best fit of P99, $%
(\Omega _{m},\Omega _{\Lambda })=(0.7,1.3)$, while for $\beta =0.5$ we
obtain $(\Omega _{m},\Omega _{\Lambda })=(0.8,0.9)$. For a flat universe,
the best fit goes from $\Omega _{m}=0.3$ when $\beta =0$ to $\Omega _{m}=0.5$
for $\beta =0.5$. While a zero-$\Lambda $ flat universe is still very
unlikely even for $\beta =0.5$, now a zero-$\Lambda $ open universe with $%
\Omega _{m}<0.5$ is within the 90\% c.l., with $\Omega _{m}=0.2$ the
best-fit zero-$\Lambda $ universe. Already for $\beta =0.2$, an open $%
\Lambda =0$ model with enough matter to account for the cluster masses, $%
\Omega _{m}=0.3$, is within the 90\% c.l.

In Fig. 4 we try instead to estimate the maximum likelihood value for $\beta 
$, constraining the universe to be flat, $\Omega _{\Lambda }=1-\Omega _{m}.$
The confidence regions are elongated along the degeneracy line $\beta
_{d}\simeq 2(\Omega _{m}-0.3)$. Here we see that to reconcile SNe Ia with
flatness and $\Lambda =0$ one needs $\beta >1$, which seems quite difficult
to achieve. The best estimate is $\left( \beta ,\Omega _{m}\right) =\left(
0.50,0.54\right) $ but the maximum is extremely flat along the degeneracy
line. Imposing both the constraint of a flat universe and $\Omega _{m}>0.2$,
we find that $\beta $ has to be larger than -0.4 (90\% c.l.).

\section{Conclusions}

The inclusion of time dependence in the fundamental constants that determine
the SNe Ia peak luminosity can modify in a significant way the best-fit
cosmological model. Here we considered the gravitational constant time
dependence, but other fundamental constants can be used as well. Modeling
both the time dependence of $G$ and the peak luminosity dependence on $G$ as
power laws, the effect on the SNe peak magnitude is proportional to the
product $\beta $ of the power law exponents. The maximum allowed value of $%
\beta $ is not very well constrained, since it depends on the precise
mechanism that powers SNe Ia. Moreover, an oscillating function $G(t)$ could
escape the local constraints.

We can summarize our results as follows:

\begin{itemize}
\item  in a flat universe, $\beta >-0.4$ if $\Omega _{m}>0.2$.

\item  for $|\beta |<0.2$ there is no significant departure from the
standard results;

\item  for $\beta \geq 0.2$ , the P99 confidence regions move by 1$\sigma $
or more toward smaller $\Lambda $, and a zero-$\Lambda $ open universe with $%
\Omega _{m}=0.3$ is within the 90\% c.l.

\item  for $\beta \geq 0.5,$ a zero-$\Lambda $ open universe with $\Omega
_{m}=0.3$ is within the 67\% c.l. region,

\item  for $\beta \geq 1$, a flat cosmology without cosmological constant
would be reconciled with the SNe Ia observations.
\end{itemize}

\vspace{1.0in}

We acknowledge useful discussions with Amedeo Tornamb\`{e}. L. A. also
thanks Emanuela Previtera for assistance in the data analysis.

\newpage

\section{Figures}

\begin{figure}
\epsfxsize 5in
\epsfbox{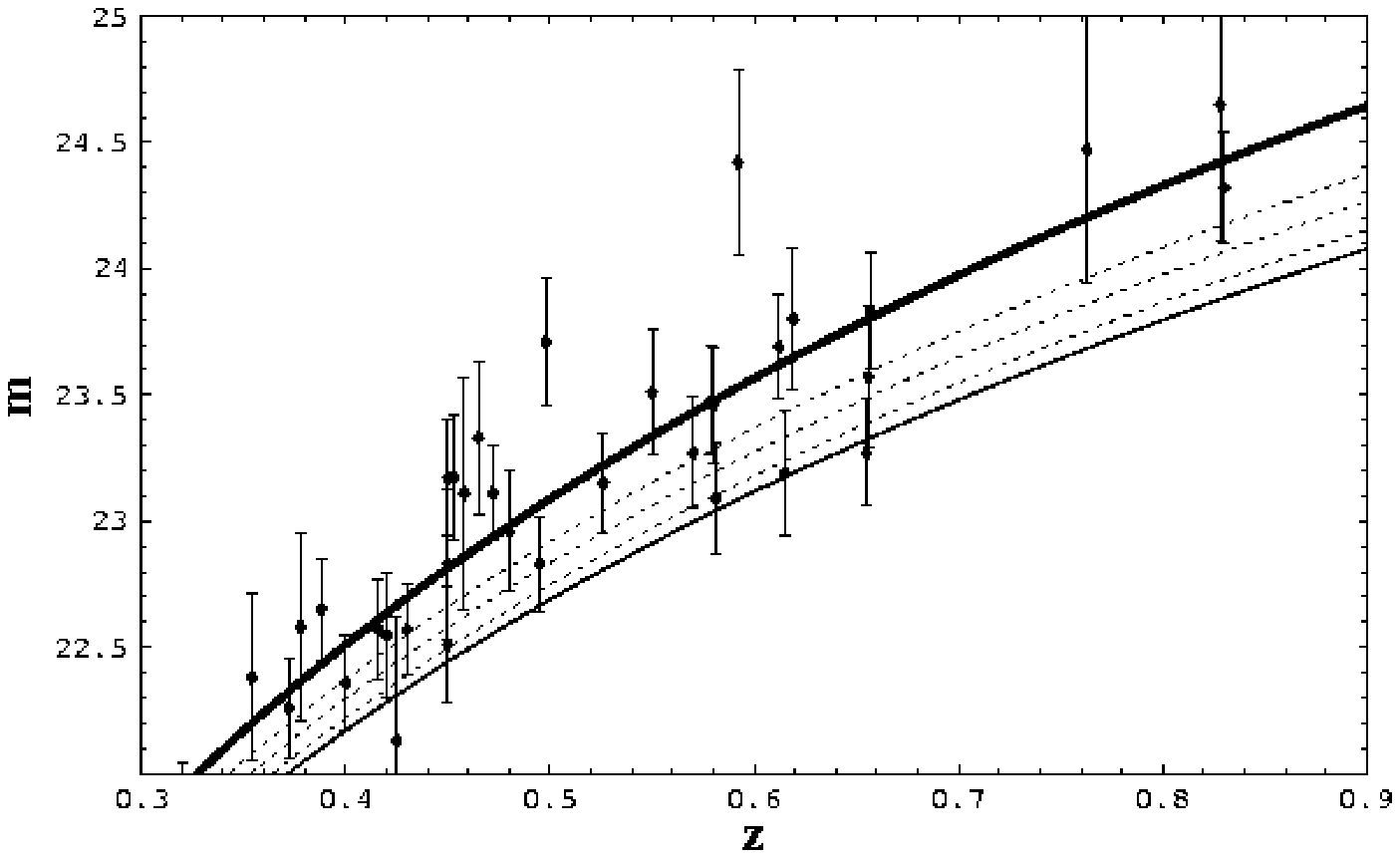}
\caption{Hubble diagram for the high-redshift SNe of P99. The thick
line is the best fit flat universe solution $(\Omega _m,\Omega _\Lambda
)=(0.28,0.72)$. The thin line corresponds to $(\Omega _m,\Omega _\Lambda
)=(1,0)$. The dotted lines are again for $(\Omega _m,\Omega _\Lambda )=(1,0)$
but with increasing $G$-correction, $\beta =0.1,0.3,0.5$, bottom to top.}
\end{figure}

\newpage

\begin{figure}
\epsfysize 4.in
\epsfbox{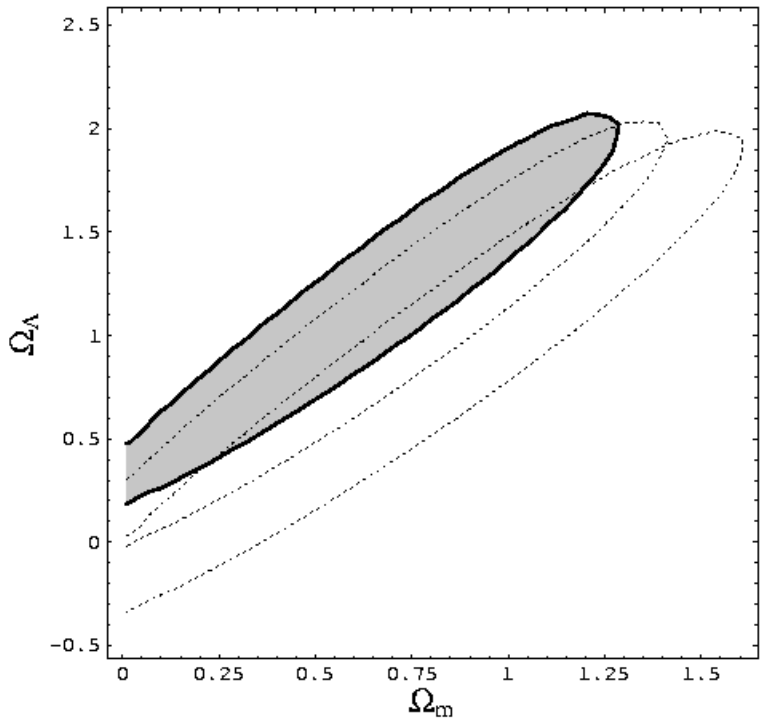}
\caption{ One sigma (67\%) confidence regions in the plane $(\Omega
_{m},\Omega _{\Lambda })$ for $\beta =0$ (shaded region) and, decreasing
toward smaller $\Omega _{\Lambda }$, for $\beta =0.2$ and $\beta =0.5$.}
\end{figure}

\begin{figure}
\epsfysize 4.in
\epsfbox{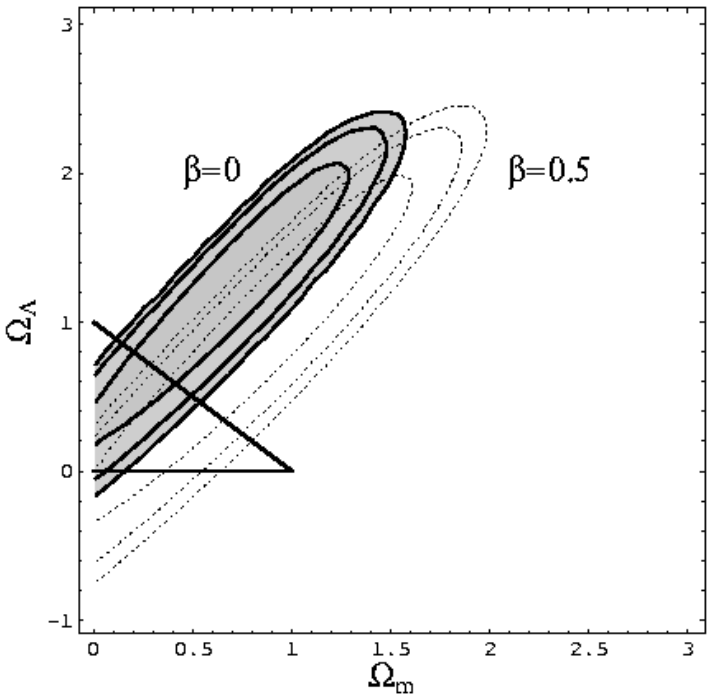}
\caption{Confidence regions (67, 90 and 95\%) in the plane $(\Omega
_{m},\Omega _{\Lambda })$ without $G$-correction (shaded region), and with a 
$\beta =0.5$ $G$-correction. The straight lines correspond to a flat
universe and a $\Lambda =0$ universe.}
\end{figure}

\begin{figure}
\epsfysize 5in
\epsfbox{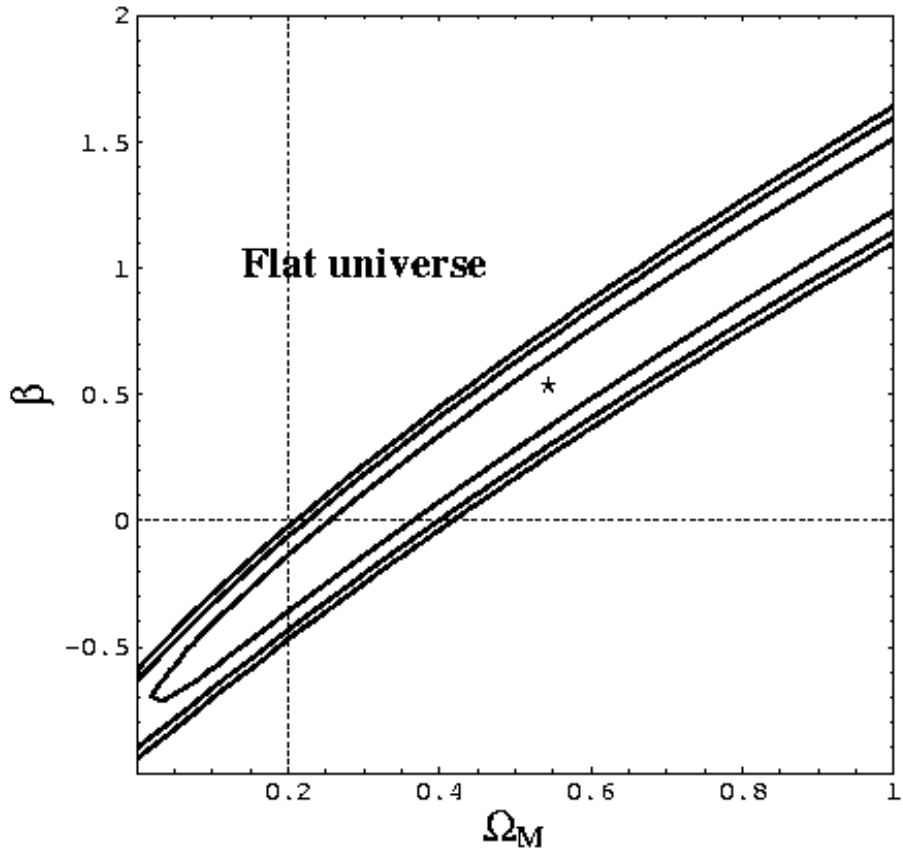}
\caption{ Confidence regions in the plane $(\beta ,\Omega _{m})$, fixing $%
\Omega _{\Lambda }=1-\Omega _{m}.$ The vertical dotted line is the minimum
value of $\Omega _{m}$ that is allowed observationally. The horizontal
dotted line is the zero $G$-correction case. The star marks the best fit
parameters.}\end{figure}

\end{document}